\documentclass[11pt,a4paper,twoside]{article}
\usepackage{graphics, graphicx}          
\usepackage{natbib}
\usepackage{color}

\newcommand{\ii}{\mathrm{i}}                                   
\newcommand{\dd}{\mathrm{d}}                                   
\newcommand{\ee}{\mathrm{e}}                                   
\newcommand{\cc}{\mathrm{cc}}                                  
\newcommand{\vc}[1]{\mbox{\protect\boldmath$#1$}}              
\newcommand{\der}[2]{\frac{\,\dd #1}{\,\dd #2}}                
\newcommand{\pder}[2]{\frac{\partial #1}{\partial #2}}         
\newcommand{\DDD}[2]{\frac{\mathrm{D}^2#1}{\mathrm{D}#2^2}}    
\newcommand{\ddd}[2]{\frac{\,\dd^2 #1}{\,\dd #2^2}}            

\fussy
\bibpunct{(}{)}{;}{a}{}{;}

\usepackage{fancyhdr}
\begin{document}
\begin{center}
  \textbf{\Large{Local Analysis of Nonlinear Oscillations of Thin Accretion Disks}}
  
\vspace{1 cm}

  \large{\textsc{Sara Fogelstr\"{o}m, Lina Levin, Rikard Slapak}}\\
  \large{\textit{Department of Physics, University of Gothenburg}}
\end{center}
\thispagestyle{plain}

\begin{abstract}
We calculated the coupling coefficients for non-linear, quasi-local oscillatory modes of thin accretion disks. We found that several of them are non-zero. Mode coupling is a necessary condition for a resonance, and thus our results may be relevant for the recently discussed QPO resonance model.
\end{abstract}

\section{Introduction}
\label{intro}
 Several galactic black hole and neutron star sources in the low mass X-ray binaries show quasi periodic variability in their observed X-ray fluxes. Some of the quasi periodic oscillations (QPOs) have high frequencies, roughly equal to orbital frequencies a few gravitational radii ($r_G = 2GM/c^2$) from the central object (with the mass $M$). The high frequency QPOs often come in pairs $(\nu_{\rm U}, \nu_{\rm L})$ of {\it twin peaks} in the Fourier power spectra (see e.g. \cite{kli00}).
Low frequency QPOs, with frequencies of a few Hz, are also observed.

There is no general agreement on a physical mechanism exciting the twin peak QPOs. \citet{abrklu01} suggested that they may be due to a resonance in some accretion disk oscillation modes. This suggestion is further supported by other observational properties of twin peak QPOs. The most important support comes from:

\begin{enumerate}
\item{\it Twin peak QPOs are oscillations (waves)}. In black holes and neutron stars, a linear correlation  was found between the high and low frequency QPOs (\cite{psa99}). The correlation proves that the QPO phenomenon is due to accretion disk oscillations, and not to kinematic effects like e.g. Doppler modulation of fluxes from isolated hot
spots: thus, QPOs are ``waves'', not ``particles''.
\item{\it Twin peak QPOs origin is connected to strong gravity}. The frequencies of twin peak kHz QPOs in microquasars scale with mass, $\nu \sim 1/M\,$, (\cite{mccrem03}). This suggests a relativistic origin of them. Indeed, in a strong gravitational field the frequency scales as $\nu \sim c/2GM/c^2 \sim 1/M\,$.
\item{\it Twin peak QPOs are due to a resonance.} In all four microquasars in which the twin peak kHz QPOs have been observed, $\nu_{\rm U}/\nu_{\rm L} = 3/2$. The commensurable frequencies obviously suggest a resonance (\cite{abrklu01}).
\end{enumerate}

The relevant ``QPO resonance pair'' should be a combination of the normal modes of the black hole accretion flow oscillations. All normal modes are known in the two mathematically idealized situations. The first one is the {\it thin disk} linear diskoseismology; see reviews by \citet{kato01} and \citet{wag99}. The second one is the {\it thick disk} slender torus linear diskoseismology, see \citet{bla06}.

Although picking up the right pair still involves guess work, several authors independently argued that the {\it epicyclic modes} or their combinations, may be particularly interesting in the context of the high frequency twin peak QPOs --- e.g. \citet{ste99}, \citet{psa00}, \citet{klu02}, \citet{kato03}, \citet{tit02}.

\citet{abr06}, \citet{bla06} and \citet{bla07} have studied properties of oscillation modes in slender tori, while Kato (see \citet{kato03}, \citet{kato04b} and \citet{kato0709}) studied a nonlinear resonant coupling between disk oscillations and its deformation. In this paper we study non-linear three-mode coupling of epicyclic and other modes in the thin disk case, using a quasi-local approach. We shall discuss similarities and differences between our and Kato's work in the last section of this article.

\section{Local Analysis of Thin Accretion Disk in Equilibrium}
By rescaling each physical quantity describing an accretion disk in equilibrium with its typical value $Q_\star$ we get $\bar{Q}=Q/Q_\star\sim1$.
In a thin accretion disk all quantities describing stationary states change quickly in the vertical direction ($z$) and slowly in the radial direction ($r$). That is
\begin{equation}
    \frac{\dd Q}{\dd r}\ll \frac{\dd Q}{\dd z}.
\end{equation}
By rescaling $r$ and $z$ by their typical values $r_\star$ and $z_\star$ so that $\bar{r}=r/r_\star\sim1$ and $\bar{z}=z/z_\star\sim1$ we write
\begin{equation}
    \frac{\dd Q}{\dd r}=\frac{Q_\star}{r_\star}\frac{\dd \bar{Q}}{\dd \bar{r}}
\end{equation}   
\begin{equation}         
    \frac{\dd Q}{\dd z}=\frac{Q_\star}{z_\star}\frac{\dd \bar{Q}}{\dd \bar{z}}
\end{equation}   
where $\dd \bar{Q}/\dd \bar{r}\sim1$ and $\dd \bar{Q}/\dd \bar{z}\sim1$. One introduces the thinness parameter $\beta\ll1$,
\begin{equation}
\Big(\frac{\dd Q}{\dd r}\Big)/\Big(\frac{\dd Q}{\dd z}\Big)= \frac{z_\star}{r_\star}\Big(\frac{\dd \bar{Q}}{\dd \bar{r}}\Big)/\Big(\frac{\dd \bar{Q}}{\dd \bar{z}}\Big)\sim\frac{z_\star}{r_\star}\equiv\beta.
\end{equation}

Considering only a small annulus of the disk, i.e. doing a local approach, it is convenient to introduce new variables $\bar{x}$ and $\bar{y}$ defined as

\begin{equation}
  \frac{r - r_0}{r_0} = \beta\bar{x},
  \quad
  \frac{z}{h_0} = \bar{y}.
  \label{eq:rescaling of r and z}
\end{equation} 

We assume the radial and vertical sizes of the ring to be comparable, that is $\ell\sim z_\star$ or $(\ell/r_0)\equiv\beta\bar{\ell}$ with $\bar{\ell}=\mathcal{O}(1)$. The part of the disk we are interested in corresponds to the ranges $-\bar{\ell}\leq\bar{x}\leq\bar{\ell}$ and $-1\leq\bar{y}\leq 1$.
 Note that the radius $r_0$ and the disk half-thickness $h(r_0)=h_0$ can be taken as typical scales ($r_\star\equiv r_0$, $z_\star\equiv h(r_0)=h_0$) so that $\beta\equiv h_0/r_0$. 

In our calculations we only consider terms of the lowest order of $\beta$. The stationary quantities which will turn out to be important are the angular velocity $\Omega=\Omega_K$ and the speed of sound $c_{s}$.

According to \citet{klu00}
\begin{equation}
n\frac{\partial c_s^2}{\partial z}=-\frac{z}{r^3},
\label{eq:kluc}
\end{equation}
 where n comes from the equation of state $p\propto\rho^\gamma=\rho^{1+1/n}$. By using the boundary condition $c_s(z=h_0)=0$ and by integrating equation (\ref{eq:kluc}) it is easy to show that
\begin{equation}
    c_s=\sqrt{\frac{h_{0}^2-z^2}{2nr_{0}^3}}.
\end{equation}
A typical scale for the speed of sound is $c_{s_\star}\equiv\omega_z(r_0) h_0/\sqrt{2n}$ so that
\begin{equation}
    \frac{c_s}{c_{s_\star}}=\sqrt{\frac{h_{0}^2-z^2}{2nr_{0}^3}}\frac{\sqrt{2n}}{\omega_{z_0}h_0}=\sqrt{\frac{h_{0}^2-z^2}{r_{0}^3\omega_{z_0}^2 h_{0}^2}}.
    \label{eq:cs1}
\end{equation}
Observing that $\omega_{z_0}^2=\omega_{z}^2 (r_0)=\Omega_{K}^2 (r_0)=r_{0}^{-3}$ equation (\ref{eq:cs1}) becomes
\begin{equation}
    \frac{c_s}{c_{s_\star}}=\sqrt{\frac{h_{0}^2-z^2}{h_{0}^2}}=\sqrt{1-\frac{z^2}{h_{0}^2}}=\sqrt{1-\bar{y}^2}.
\end{equation}

\section{Perturbation Equations}

Next thing to do is to perturb the equation of motion. This is done by the Lagrangian approach also used by \citet{now91}, that is 
\begin{equation}
  \Delta\left(\frac{\mathrm{D}\vc{v}}{\mathrm{D}t} + \frac{1}{\rho}\nabla p
  +\nabla \Phi - \frac{1}{\rho}\vc{F}_\mathrm{v}\right)=0
\end{equation}
where $p$ is pressure, $\Phi$ is gravitational potential and $\vc{F}_v$ is viscous forces. The Lagrangian perturbation is defined as $\Delta Q(x,t)=Q(x+\xi,t)-Q_0(x,t)$ where $\xi$ is called the Lagrangian displacement. Considering a polytropic fluid the perturbation becomes
\begin{eqnarray}
\DDD{\xi^i}{t} - \frac{1}{\rho}\nabla_j\left[p(\gamma-1)(\nabla\cdot\vc{\xi})g^{ij} + 
  p\nabla^i\xi^j\right] + \xi^j\nabla_j\nabla^i\Phi = {a}^i_\mathrm{v}(\vc{\xi})  +{a}^i_\mathrm{n}(\vc{\xi})
  \label{eq:pert}
\end{eqnarray}
where $g^{ij}$ is the metric tensor and $\mathrm{D}/\mathrm{D}t\equiv\partial/\partial t
+v^k\nabla_k$ is the Lagrangian flow derivative, see e.g. \citet{frisch78} and \citet{lynost67}. The nonlinear terms are collected in $\vc{a}_\mathrm{n}(\vc{\xi})$, and $\vc{a}_\mathrm{v}(\vc{\xi})$ is the contribution of the viscous forces which are ignored in the calculations. When using cylindrical coordinates $(r,\phi,z)$ equation (\ref{eq:pert}) becomes

\begin{eqnarray}
 \!\!\!&\phantom{=}&\!\!\! \ddd{\xi^{r}}{t} 
  -2\Omega\der{\xi^{\phi}}{t} 
  -\left[\Omega^2-\frac{\partial^2 \Phi}{\partial r^2}\right]\xi^{r}
  +\xi^{\phi}(\vc{v}_\mathrm{p}\cdot\nabla)\Omega
     +\frac{\partial^2\Phi}{\partial r\partial z}\xi^{z}
\nonumber\\ \!\!\!&\phantom{=}&\!\!\! 
  -(\nabla\cdot\vc{\xi})\pder{c_\mathrm{s}^2}{r}
  -c_\mathrm{s}^2\pder{(\nabla\cdot\vc{\xi})}{r}
  -n\pder{c_\mathrm{s}^2}{r}\pder{\xi^{r}}{r}
   -n\pder{c_\mathrm{s}^2}{z}\pder{\xi^{z}}{r}
   ={a}^r_\mathrm{v}(\vc{\xi}) 
  +{a}^r_\mathrm{n}(\vc{\xi}),
  \label{eq:pert_r}
\end{eqnarray}  

\begin{eqnarray}  
\!\!\!&\phantom{=}&\!\!\!    \ddd{\xi^{\phi}}{t}
  +2\Omega\der{\xi^{r}}{t} 
  -\left[\Omega^2 - \frac{1}{r}\pder{\Phi}{r}\right]\xi^{\phi}
  +\xi^{r}(\vc{v}_\mathrm{p}\cdot\nabla)\Omega
-\frac{v^r}{r}\left[\Omega\xi^{r} + \frac{v^r}{r}\xi^{\phi}\right]
\nonumber\\
\!\!\!&\phantom{=}&\!\!\!  
    -\frac{1}{r}\left[c_\mathrm{s}^2\pder{(\nabla\cdot\vc{\xi})}{\phi}
  +n\pder{c_\mathrm{s}^2}{r}\pder{\xi^{r}}{\phi}\right.
  \left.+n\pder{c_\mathrm{s}^2}{z}\pder{\xi^{z}}{\phi}\right]
     ={a}^\phi_\mathrm{v}(\vc{\xi}) 
  +{a}^\phi_\mathrm{n}(\vc{\xi}),
  \label{eq:pert_phi}
\end{eqnarray}

\begin{eqnarray}
\!\!\!&\phantom{=}&\!\!\!  \ddd{\xi^{z}}{t} + \frac{\partial^2\Phi}{\partial r\partial z}\xi^{r} +
  \frac{\partial^2\Phi}{\partial z^2}\xi^{z}-
    (\nabla\cdot\vc{\xi})\pder{c_\mathrm{s}^2}{z} 
    -c_\mathrm{s}^2\pder{(\nabla\cdot\vc{\xi})}{z}
  \nonumber\\
\!\!\!&\phantom{=}&\!\!\!
  -n\pder{c_\mathrm{s}^2}{r}\pder{\xi^{r}}{z}-
  n\pder{c_\mathrm{s}^2}{z}\pder{\xi^{z}}{z}
  ={a}^z_\mathrm{v}(\vc{\xi}) 
  +{a}^z_\mathrm{n}(\vc{\xi})
  \label{eq:pert_z}
\end{eqnarray}
where the operators are defined as 
\begin{equation}
  \der{}{t}\equiv \pder{}{t} + v^k\pder{}{x^k},
  \quad
  (\vc{v}_\mathrm{p}\cdot\nabla)\equiv v^r\pder{}{r} + v^z\pder{}{z}.
\end{equation}
The stationary quantities appearing in these equations have been calculated. In addition expressions for the gravitational terms are needed. Some short and simple calculations yield

\begin{eqnarray}
\!\!\!&\phantom{=}&\!\!\!
    \left[\frac{\partial^2 \Phi}{\partial z^2}\right]_{z=0}=\omega_z^2,
    \quad
    \left[\frac{\partial^2 \Phi}{\partial r^2}\right]_{z=0}=\omega_r^2 -3\Omega^2,
    \nonumber\\
\!\!\!&\phantom{=}&\!\!\!    \left[\frac{\partial^2 \Phi}{\partial r \partial z}\right]_{z=0}=0,
    \quad
    \left[\frac{1}{r}\pder{\Phi}{r}\right]_{z=0}=\omega_r^2.
\end{eqnarray}
Equations (\ref{eq:pert_r}) - (\ref{eq:pert_z}) are now divided by $\Omega_{\star}^2$. When collecting the zeroth order terms in $\beta$ the resulting equations become
\begin{eqnarray}
\der{\xi^{\phi}}{\bar{t}}+2\xi^{r} = 0,
  \label{eq:pert0_phi}
\\
\frac{\dd^2\xi^{r}}{\dd\bar{t}^2} + \bar{\omega}_r^2\xi^{r} + 
  \bar{\omega}_z^2\left[\bar{y}\pder{\xi^{z}}{\bar{x}}-
  \frac{1}{2n}(1-\bar{y}^2)\pder{\varphi}{\bar{x}}\right] = 0,
  \label{eq:pert0_r}
\\
\frac{\dd^2\xi^{z}}{\dd\bar{t}^2} + \bar{\omega}_z^2\xi^{z} +
  \bar{\omega}_z^2\left[
  \frac{1}{n}\bar{y}\varphi - \frac{1}{2n}(1-\bar{y}^2)\pder{\varphi}{\bar{y}} +
  \bar{y}\pder{\xi^{z}}{\bar{y}}\right]   = 0,
  \label{eq:pert0_z}
\end{eqnarray}
where $\bar{t}\equiv\Omega_\star t$. This is the same result as was derived by \citet{now91}. At this level of approximation $\dd/\dd\bar{t}=\partial/\partial\bar{t}+\partial/\partial\phi$ and the variable $\varphi$ has been introduced as
\begin{equation}
  \varphi\equiv\pder{\xi^{r}}{\bar{x}} + \pder{\xi^{z}}{\bar{y}}.
\end{equation}

\section{Lowest-order Linear Equations}

The coefficients in the linear equations (\ref{eq:pert0_r}) and (\ref{eq:pert0_z}) do not depend explicitly on $\bar{t}$, $\phi$ and $\bar{x}$. Therefore, the solutions of these equations can be found on the form

\begin{eqnarray}
  \xi^{r}=\ee^{-\ii(\bar{\omega}\bar{t} + m\phi + \bar{k}\bar{x})} f(\bar{y}),
  \nonumber\\
  \xi^{z}=\ee^{-\ii(\bar{\omega}\bar{t} + m\phi + \bar{k}\bar{x})} g(\bar{y}),
  \label{eq:xirxiz}
\end{eqnarray}
where $\bar{k}$ is a scaled component of the radial wave-vector.
Periodic boundary conditions $\bar{k}=\pi n_r/\bar{\ell}$ are assumed, with $n_r$ being an integer wave-number. The wave vector in physical units is given by $k=\bar{k}h^{-1}=\pi n_r/\ell$. Using the expressions in (\ref{eq:xirxiz}) equation (\ref{eq:pert0_r}) yields

\begin{equation}
  f=\ii\bar{k}\bar{\omega}_z^2\frac{2n\bar{y}g - (1-\bar{y}^2)g^\prime}
  {2n(\bar{\sigma}^2-\bar{\omega}_r^2)-k^2\omega_z^2(1-\bar{y}^2)}.
  \label{eq:f}
\end{equation}
When substituting this result into equation (\ref{eq:pert0_z}) a linear second-order differential equation for $g(\bar{y})$ is obtained
\begin{equation}
  (1-\bar{y}^2)g^{\prime\prime} - \bar{y}U(\bar{y})g^\prime + V(\bar{y})g = 0
  \label{eq:g}
\end{equation}
where the coefficients are given by
\begin{eqnarray}
  U(\bar{y}) &\equiv& 2n\,\frac{2(n+1)(\bar{\sigma}^2-\bar{\omega}_r^2)-\bar{k}^2\omega_z^2(1-\bar{y}^2)}
  {2n(\bar{\sigma}^2-\bar{\omega}_r^2)-\bar{k}^2\omega_z^2(1-\bar{y}^2)},
  \label{eq:B1}
\end{eqnarray}

\begin{eqnarray}
V(\bar{y}) \equiv
\frac{2n(\bar{\sigma}^2-\bar{\omega}_r^2)(\bar{\sigma}^2-\bar{\omega}_z^2)-
  \bar{\sigma}^2\bar{\omega}_z^2 \bar{k}^2(1-\bar{y}^2)}{(\bar{\sigma}^2-\bar{\omega}_r^2)\omega_z^2}
  +\frac{4n\omega_z^2 \bar{k}^2 \bar{y}^2}{2n(\bar{\sigma}^2-\bar{\omega}_r^2)-\bar{k}^2\omega_z^2(1-\bar{y}^2)}.
\label{eq:C1}
\end{eqnarray}

Equation (\ref{eq:g}) must be completed by the boundary condition stating that 
the Lagrangian pressure variations $\Delta p$ vanishes at the surface
of the unperturbed disk. Since
\begin{equation}
  \Delta p = - \left(1+\frac{1}{n}\right)p\,\nabla_i\xi^i
\end{equation}
and the vertical profile of the pressure is given by $p\propto(1-\bar{y}^2)^{n+1}$, this
condition takes the form
\begin{eqnarray}
\!\!\!&\phantom{=}&\!\!\!  \lim_{\bar{y}\rightarrow\pm 1}\varphi(\bar{y})\left(1-\bar{y}^2\right)^{n+1}=0,
  \nonumber \\
\!\!\!&\phantom{=}&\!\!\!  \varphi(\bar{y}) = 
  \frac{2n(\bar{\sigma}^2-\bar{\omega}_r^2)g^\prime - \bar{k}^2\omega_z^2\bar{y}^2 g}{
  2n(\bar{\sigma}^2-\bar{\omega}_r^2)-\bar{k}^2\omega_z^2(1-\bar{y}^2)}.
  \label{eq:bound1}
\end{eqnarray}
Other boundary conditions are set in the equatorial plane. Considering
either even or odd modes the two possible conditions imposed on $g$ are
\begin{equation}
  g(0) = 0, 
  \quad\mathrm{or}\quad
  g^\prime(0)=0.
  \label{eq:bound2}
\end{equation}

\subsection{Isothermal Disk}
It is possible to get the equations for vertically perturbed isothermal disk from (\ref{eq:g}) if $n\rightarrow\infty$. But this limit cannot be applied directly to equation (\ref{eq:g}), because it is derived under the condition $h\ll r$. Here the disk thickness, $h_0$, is defined as the height in the z-direction where the speed of sound vanishes, i.e. $c_s(z=h_0)=0$. The sound speed is constant in vertically isothermal disks. This leads to that the thickness goes to infinity due to $h\propto n^{1/2}$. Therefore the disk thickness must be redefined. This is done with a new thickness, $h^\star$, assuming it remains small and approximately constant when $n\rightarrow\infty$. The two thicknesses are related as $h^\star = h/(2q n)^{1/2}$, where $q$ is a dimensionless constant. With the new definition of the disk thickness the scaled lengths $\bar{x}$ and $\bar{y}$ also need to be redefined. The new variables $\bar{x}^\star$ and $\bar{y}^\star$ are related to the old ones as $\bar{x}^\star\equiv(2q n)^{1/2}\bar{x}$ and $\bar{y}^\star\equiv(2q n)^{1/2}\bar{y}$. The boundary conditions are the same as before. The radial wave-vector $\bar{k}^\star\equiv\bar{k}/(2q n)^{1/2}$ has to be reintroduced as well. Now equation (\ref{eq:g}) becomes
 
\begin{eqnarray}
\left(2q n-\bar{y}^{\star2}\right) \frac{\dd^2 g}{\dd \bar{y}^{\star 2}} - 
  \bar{y}^\star U\left(\frac{\bar{y}^\star}{\sqrt{2q n}}\right) 
  \frac{\dd g}{\dd \bar{y}^\star} +
V\left(\frac{\bar{y}^\star}{\sqrt{2q n}}\right) g = 0
\end{eqnarray}
which in the limit $n\rightarrow\infty$ gives
\begin{equation}
  q\,g^{\prime\prime} - \bar{y} g^\prime + 
  \frac{(\bar{\sigma}^2-\bar{\omega}_r^2)(\bar{\sigma}^2-\bar{\omega}_z^2)-
  q\bar{\sigma}^2\bar{\omega}_z^2 \bar{k}^2}{(\bar{\sigma}^2-\bar{\omega}_r^2)\bar{\omega}_z^2}    \,g = 0,
  \label{eq:g_iso1}
\end{equation}
where the stars in $\bar{y}$ and $\bar{k}$ were omitted. The meaning of $q$ is apparent from the following relation valid for $n<\infty$
\begin{equation}
  q = \frac{1}{2n}\left(\frac{h}{h^\star}\right)^2 = \left(\frac{c_\mathrm{s}}{\omega_z h}\right)^2.
\end{equation}
This can be used to rewrite equation (\ref{eq:g_iso1}) as (as was done by \citet{oka87} and by Kato in e.g. \citet{kato04b})
\begin{eqnarray}
\!\!\!&\phantom{=}&\!\!\!  q\,g^{\prime\prime} - \bar{y} g^\prime + 
  (n_z - 1)g = 0,
  \nonumber \\
\!\!\!&\phantom{=}&\!\!\!   n_z = \frac{(\sigma^2 - \omega_r^2 - c_\mathrm{s}^2 k^2)\sigma^2}
  {(\sigma^2-\omega_r^2)\omega_z^2}.
  \label{eq:g_iso2}
\end{eqnarray}
Note that the case $n_z=0$ is excluded from the discussion because it violates
equation (\ref{eq:f}). This case corresponds to purely horizontal oscillations. 

The factor $q$ in equation (\ref{eq:g_iso2}) can be eliminated by choosing it to be equal to one. Then equation (\ref{eq:g_iso2}) will become a Hermite equation. The boundary conditions for this equation is (\cite{oka87})

\begin{equation}
  \lim_{\bar{y}\rightarrow\pm\infty} \ee^{-\bar{y}^2/4} g(\bar{y}) = 0,
  \label{eq:bound_iso1}
\end{equation}
which replaces condition (\ref{eq:bound1}). The conditions (\ref{eq:bound2}) remain valid in the case of the isothermal disk.

The solution to equation (\ref{eq:g_iso2}) that satisfies the boundary conditions exists only when $n_z$  takes a positive integer value. Then the solutions are given by Hermite polynomials. Now that $g$ is known it is possible to get $f$ as well and then the Lagrangian displacement $\vc{\xi}(\bar{x},\bar{y},\phi)$ is given by (time-dependence is omitted) 

\begin{eqnarray}
  \xi^{r} &=& \ii K \frac{\bar{k}}{n_z}\frac{\bar{\sigma}^2}{\bar{\sigma}^2-\bar{\omega}_r^2}\,\,
  H\left(n_z,{\bar{y}}\right) 
  \ee^{-\ii(m\phi + \bar{k}\bar{x})},
  \label{eq:xi_r}
  \\
  \xi^{\phi} &=& 2 K\frac{\bar{k}}{n_z}\frac{\bar{\sigma}}{\bar{\sigma}^2-\bar{\omega}_r^2}\,\,
  H\left(n_z,{\bar{y}}\right) 
  \ee^{-\ii(m\phi + \bar{k}\bar{x})},
  \\
  \xi^{z} &=& K\,H\left(n_z\!-\!1,{\bar{y}}\right)\,\ee^{-\ii(m\phi + \bar{k}\bar{x})},
  \label{eq:xi_z}
\end{eqnarray}
where $K$ is a normalization constant, and $n_z = 1,2,3\dots$. Hence, a linear mode is determined by three natural numbers: $m$, $n_r$ and $n_z$. For a particular mode $n_z-1$ describes the number of modes in the vertical direction. When $n_z$ is even/odd the corresponding mode is symmetric/antisymmetric under the reflection about equatorial plane. The modes with $(n_r,n_z)=(0,1)$ and $(1,2)$ are analogical to the vertical epicyclic and breathing mode of the slender torus, respectively.

\subsection{Purely Vertical Oscillations}
Instead of assuming an isothermal disk, i.e. here $\gamma\neq1$, equation (\ref{eq:g}) can be made easier to solve by putting $k=0$. Then only vertical oscillations will survive. This implies that equation (\ref{eq:B1}) and equation (\ref{eq:C1}) become
\begin{equation}
  U(\bar{y}) = 2(n+1),
\end{equation}

\begin{equation}
  V(\bar{y}) = \frac{2n(\bar{\sigma}^2-\bar{\omega}_z^2)}{\bar{\omega}_z^2}.
\end{equation}
This changes equation (\ref{eq:g}) into
\begin{equation}
  (1-\bar{y}^2)g^{\prime\prime}-2(n+1)\bar{y}g^{\prime}+\frac{2n(\bar{\sigma}^2-\bar{\omega}_z^2)}{\bar{\omega}_z^2}g = 0.
\label{eq:g,k=0}
\end{equation}
The solutions to equation (\ref{eq:g,k=0}) are Jacobi polynomials on the form $g=P_N^{(n,n)}(\bar{y})$, where N=0,1,2... The first four Jacobi polynomials take the forms

\begin{eqnarray}
P_0^{(n,n)}(\bar{y})\!\!\!&=&\!\!\!1,\nonumber \\
P_1^{(n,n)}(\bar{y})\!\!\!&=&\!\!\!(n+1)\bar{y},\nonumber \\
P_2^{(n,n)}(\bar{y})\!\!\!&=&\!\!\!\frac{1}{4}(n+2)(2n+3)\bar{y}^2-\frac{1}{2}(n+2),\nonumber \\
P_3^{(n,n)}(\bar{y})\!\!\!&=&\!\!\!\frac{(n+3)(n+2)}{8} \left(\frac{4n+10}{3} \bar{y}^3-2\bar{y} \right).
\end{eqnarray}
The dispersion relation in this case becomes
\begin{equation}
  \bar{\sigma}^2=(N+1)\bar{\omega}_z^2\left(\frac{N}{2n}+1\right).
\end{equation}
The Lagrangian displacement for vertical oscillations is given by 
\begin{eqnarray}
\!\!\!&\phantom{=}&\!\!\!\
\xi^r=0
\nonumber\\
\!\!\!&\phantom{=}&\!\!\!\
\xi^{\phi}=0
\nonumber\\
\!\!\!&\phantom{=}&\!\!\!\
\xi^z=K\ee^{-im\phi}P_N^{(n,n)}(\bar{y})
\end{eqnarray}
where the time dependence is omitted.

\section{Nonlinear Pressure Coupling}
\label{sec5}
In the calculations to obtain the nonlinear pressure coupling coefficients we closely follow the formalism of \citet{sch01}. A compact way to write equations (\ref{eq:pert_r})--(\ref{eq:pert_z}) is

\begin{equation}
  \frac{\partial^2\vc{\xi}}{\partial t^2} + S \frac{\partial\vc{\xi}}{\partial t} + T \vc{\xi}(x)=0,
  \label{eq:lin0}
\end{equation}
The solution to equation (\ref{eq:lin0}) can be assumed to be on the form of normal modes
\begin{equation}
	\vc{\xi}_\alpha(x,t)= \ee^{-i \omega_\alpha t} \vc{\xi}_\alpha(x)
	\label{eq:asump}
\end{equation}
where $\vc{\xi}_\alpha$ is the corresponding eigenfunction. Inserting the assumption into equation (\ref{eq:lin0}) gives
\begin{equation}
	-{\omega_\alpha}^2 \vc{\xi}_\alpha(x) -i\omega_\alpha S \vc{\xi}_\alpha(x) + T \vc{\xi}_\alpha(x) = 0.
	\label{eq:homo}
\end{equation} 
The linear modes are determined by three integer numbers $m$, $n_r$ and $n_z$ and form a complete, generally non-orthonormal basis in the Hilbert space $\mathcal{H}$ of all possible Lagrangian displacements with the scalar product defined as
\begin{equation}
  \langle \vc{\xi},\vc{\xi}^\prime\rangle\equiv\int_V\rho\,\vc{\xi}^\star\cdot\vc{\xi}^\prime\,\dd V.
\label{eq:scalar}
\end{equation} 
Inspecting equations (\ref{eq:pert_r})--(\ref{eq:pert_z}) we find that the operators S and T are respectively anti-Hermitian and Hermitian with respect to the above scalar product. As shown by \citet{sch01} the eigenfunctions of the modes satisfy the pseudo-orthogonality relation
\begin{equation}
	(\omega_\alpha + \omega_\beta)\langle \vc{\xi}_\alpha,\vc{\xi}_\beta\rangle + \langle\vc{\xi}_\alpha,iS\vc{\xi}_\beta\rangle = 0.
\end{equation}

Looking at equation (\ref{eq:lin0}) and now including nonlinear terms in $\vc{\xi}$ but still only zeroth order terms in $\beta$ the equation will become
\begin{equation}
  \frac{\partial^2\vc{\xi}}{\partial t^2} + S \frac{\partial\vc{\xi}}{\partial t} + T\vc{\xi}(x)=\vc{a}(\vc{\xi}),
  \label{eq:nlin0}
\end{equation}
where $\vc{a}(\vc{\xi})$ represents effects of nonlinear accelerations. Thanks to the completeness of the linear-mode eigenfunctions, the solution of the nonlinear equation (\ref{eq:nlin0}) can be found on the form
\begin{equation}
  \vc{\xi}(t,\phi,\bar{x},\bar{y}) = \sum_\alpha c_\alpha(t)\vc{\xi}_\alpha(\phi,\bar{x},\bar{y}) + \cc
  \label{eq:xi_alpha}
\end{equation}
where `cc` denotes complex conjugate and $\vc{\xi}_{\alpha}$ denotes eigenfunctions of the linear modes, $\alpha=(m,n_r,n_z)$. The problem with non-orthogonality of the basis can be handled by carrying out this expansion in phase-space rather than just in configuration space (\cite{sch01}). Hence the time derivative of the Lagrangian displacement is expanded as
\begin{equation}
	\partial_t\vc{\xi}(t,\phi,\bar{x},\bar{y}) = \sum_\alpha -i\omega_\alpha c_\alpha(t)\vc{\xi}_\alpha(\phi,\bar{x},\bar{y}) + cc.
\end{equation}
 
Considering three different expressions using equation (\ref{eq:xi_alpha})
\begin{eqnarray}
	\langle\vc{\xi}_\alpha,\omega_\alpha \vc{\xi}\rangle = 
	\omega_\alpha \langle\vc{\xi}_\alpha,\sum c_\beta \vc{\xi}_\beta\rangle = 
	\sum c_\beta \omega_\alpha\langle\vc{\xi}_\alpha,\vc{\xi}_\beta\rangle,
\label{eq:b_alpha1}
\end{eqnarray}	 

\begin{equation}
	\langle\vc{\xi}_\alpha,i \vc{\dot{\xi}}\rangle = 
	\langle\vc{\xi}_\alpha,i \sum_\beta -i\omega_\beta c_\beta \vc{\xi}_\beta\rangle =
	\sum \omega_\beta c_\beta \langle\vc{\xi}_\alpha,\vc{\xi}_\beta\rangle,
\label{eq:b_alpha2}
\end{equation}

\begin{equation}
	\langle\vc{\xi}_\alpha,iS \vc{\xi}\rangle =
	\langle iS\vc{\xi}_\alpha,\sum c_\beta \vc{\xi}_\beta\rangle =
	\sum c_\beta\langle iS \vc{\xi}_\alpha,\vc{\xi}_\beta\rangle
\label{eq:b_alpha3}
\end{equation}
and then summing these gives
\begin{eqnarray}
\langle \vc{\xi}_\alpha,\omega_\alpha\vc{\xi}+i\vc{\dot{\xi}}+iS\vc{\xi}\rangle = \sum_\beta c_\beta(\langle \vc{\xi}_\alpha,\vc{\xi}_\beta\rangle(\omega_\alpha+\omega_\beta) +
	\langle iS\vc{\xi}_\alpha,\vc{\xi}_\beta\rangle).
\label{sumb_alpha}
\end{eqnarray}
Due to that non-zero terms only exist when $\alpha=\beta$ (as follows from the
pseudo-orthogonality condition) equation (\ref{sumb_alpha}) will become

\begin{equation}
	\langle \vc{\xi}_\alpha,\omega_\alpha\vc{\xi}+i\vc{\dot{\xi}}+iS\vc{\xi}\rangle =
	c_\alpha (2\omega_\alpha\langle \vc{\xi}_\alpha,\vc{\xi}_\alpha\rangle + 
	\langle \vc{\xi}_\alpha, iS\vc{\xi}_\alpha\rangle),
	\label{eq:c_alphab_alpha}
\end{equation}
where $(2\omega_\alpha\langle \vc{\xi}_\alpha,\vc{\xi}_\alpha\rangle + 
	\langle \vc{\xi}_\alpha, iS\vc{\xi}_\alpha\rangle)$ is defined as $b_\alpha$. 

Looking again at equation (\ref{eq:nlin0}) and performing a scalar product with $\vc{\xi}_\alpha$ gives
\begin{eqnarray}
	\!\!\!&\phantom{=}&\!\!\!\underbrace{\langle \vc{\xi}_\alpha,\frac{\partial\vc{\dot{\xi}}}{\partial t}+S\vc{\dot{\xi}}\rangle}_{I} +
	\underbrace{\langle \vc{\xi}_\alpha,T\vc{\xi}\rangle}_{II} = 
	\langle \vc{\xi}_\alpha,\vc{a}(\vc{\xi})\rangle.
	\label{eq:I_II}
\end{eqnarray}
Expanding I and II separately will give
\begin{eqnarray}
  I\!\!\!&=&\!\!\! \frac{\partial}{\partial t} \langle \vc{\xi}_\alpha,\vc{\dot{\xi}}+S\vc{\xi}\rangle =
	-i\frac{\partial}{\partial t}\langle \vc{\xi}_\alpha,i\vc{\dot{\xi}}+iS\vc{\xi}+\omega_\alpha \vc{\xi}\rangle+
	i\frac{\partial}{\partial t}\langle \vc{\xi}_\alpha,\omega_\alpha\vc{\xi}\rangle
	\nonumber\\ \!\!\!&=&\!\!\!	
		-i\frac{\partial}{\partial t}c_\alpha b_\alpha+
	\frac{\partial}{\partial t}\langle \vc{\xi}_\alpha,\omega_\alpha\vc{\xi}\rangle
	=-ib_\alpha \frac{\dd c_\alpha}{\dd t}+\langle \vc{\xi}_\alpha,i\omega_\alpha\vc{\dot{\xi}}\rangle,
\end{eqnarray}
\begin{eqnarray}
	II\!\!\!&=&\!\!\!\langle \vc{\xi}_\alpha,T\vc{\xi}\rangle=\langle T\vc{\xi}_\alpha,\vc{\xi}\rangle 
=\langle \omega_\alpha^2\vc{\xi}_\alpha+i\omega_\alpha S\vc{\xi}_\alpha,\vc{\xi}\rangle
\nonumber\\ \!\!\!&=&\!\!\!	
  \omega_\alpha \langle\omega_\alpha\xi_\alpha+ iS\vc{\xi}_\alpha,\vc{\xi}\rangle= 
 	\omega_\alpha \langle \vc{\xi}_\alpha,\omega_\alpha\vc{\xi}+iS\vc{\xi}\rangle,
\end{eqnarray}
where equations (\ref{eq:c_alphab_alpha}) and (\ref{eq:homo}) have been used. 
Equation (\ref{eq:I_II}) now becomes
\begin{eqnarray}
	-ib_\alpha \frac{\dd c_\alpha}{\dd t}+\omega_\alpha \langle \vc{\xi}_\alpha,i\vc{\dot{\xi}}+
	\omega_\alpha\vc{\xi}+iS\vc{\xi}\rangle=\langle \vc{\xi}_\alpha,\vc{a}(\vc{\xi})\rangle.
	\label{eq:I_II2}
\end{eqnarray}
Rearranging equation (\ref{eq:I_II2}) gives
\begin{equation}
  \der{c_\alpha}{t} + \ii\omega_\alpha c_\alpha = \frac{\ii}{b_\alpha}
  \langle{\vc{\xi}}_\alpha,\vc{a}(\vc{\xi})\rangle,
  \label{eq:nlin1}
\end{equation}
which can be seen as a set of ordinary differential equations for many harmonic oscillators. The scalar product on the right-hand side of equation (\ref{eq:nlin1}) can be expressed as

\begin{eqnarray}
 \langle{\vc{\xi}}_\alpha,\vc{a}(\vc{\xi})\rangle =  \sum_{\beta,\gamma}
  \kappa_{\bar{\alpha}\beta\gamma}c_\beta c_\gamma +
  \kappa_{\bar{\alpha}\bar{\beta}\gamma}c_\beta^\star c_\gamma +
  \kappa_{\bar{\alpha}\beta\bar{\gamma}}c_\beta c_\gamma^\star +
  \kappa_{\bar{\alpha}\bar{\beta}\bar{\gamma}}c_\beta^\star c_\gamma^\star,
\end{eqnarray}
when considering only quadratic terms of $\vc{\xi}$ in $\vc{a}(\vc{\xi})$ and where $\kappa$ is defined as
\begin{equation}
	\kappa_{\alpha\beta\gamma}=\langle \vc{\xi}_\alpha,\vc{a}(\vc{\xi}_\beta,\vc{\xi}_\gamma)\rangle.
\end{equation}

In order to get the coupling coefficients it is necessary to find an explicit expression for the acceleration. It can be done by adopting the formalism used by \citet{kumgol89}, starting with expanding the Lagrangian density as
\begin{equation}
	L = L_0+L_1+L_2+L_3+\ldots,
	\quad\quad L_n\propto \left|\vc{\xi}\right|^n,
\end{equation}
where $L_1$ is the first order contribution describing the stationary flow while the second order part, $L_2$, gives the equation for linear perturbations. The higher order terms gives the nonlinear evolution of the linear modes. Expanding $L$ leads to that the Euler-Lagrangian equations can be applied as follows
\begin{eqnarray}
EL(L)=EL(L_0)+EL(L_1)+EL(L_2)+EL(L_3)+\ldots=0.
\end{eqnarray}
Considering only the first orders of perturbations gives
\begin{equation}
	EL(L_2)=-EL(L_3)\equiv \rho \vc{a}(\vc{\xi})
	\label{eq:rhoA}
\end{equation}
since both $EL(L_0)$ and $EL(L_1)$ are zero. $\vc{a}(\vc{\xi})$ is as before an acceleration due to the nonlinear terms. The equation describing the Lagrangian density is
\begin{eqnarray}
L=\frac{1}{2}\rho(v^{i}+\partial_t \xi^{i}+v^k \nabla_k \xi^{i})^2 - p\frac{J^{1-\gamma}}{\gamma -1} - \rho\Phi(x^{i}+\xi^{i}),
	\label{eq:L}
\end{eqnarray}
where $v^{i}, \rho$ and p are the velocity, density and pressure of the stationary flow, $\Phi$ is the gravitational potential and J is the Jacobian defined as

\begin{eqnarray}
	J^{1-\gamma}\!\!\!&=&\!\!\!1+(1-\gamma)\nabla\vc{\xi}+
	\frac{1}{2}(1-\gamma)((1-\gamma)(\nabla\vc{\xi})^2- 
		\nabla_i \xi^j\nabla_j\xi^{i})-
	\nonumber\\ \!\!\!&\phantom{=}&\!\!\!
		\frac{1}{6}(1-\gamma)((\gamma-1)^2(\nabla\vc{\xi})^3+
		3(\gamma-1)(\nabla\vc{\xi})\nabla_j\xi^k
	\nabla_k\xi^j+2\nabla_i\xi^j\nabla_j\xi^k\nabla_k\xi^{i}.
	\label{eq:Jacobi}
\end{eqnarray}
To get $L_3$ only terms with $\vc{\xi}^3$-dependence contribute. This leads to that only the second term in equation (\ref{eq:L}) contributes to $L_3$. Using equation (\ref{eq:Jacobi}) and the fact that $\vc{a}(\vc{\xi})=EL(L_3)/\rho$ (from equation (\ref{eq:rhoA})) it is now possible to write down an expression for the coupling coefficients. The three-mode coupling coefficient $\kappa_{\alpha\beta\gamma}$ is then given by
\begin{eqnarray}
  \kappa_{\alpha\beta\gamma}\!\!\!&=&\!\!\!
  \frac{1}{2}\int_V \Big\{
  p(\gamma-1)^2\eta_\alpha\eta_\beta\eta_\gamma + 3 p(\gamma-1)\eta_{[\alpha}\eta_{\beta\gamma]}+
  2p\eta_{\alpha\beta\gamma} - 
    \rho\,\xi^i_\alpha\xi^j_\beta\xi^k_\gamma\nabla_i\nabla_j\nabla_k\Phi
  \Big\}\,\dd V,
  \label{eq:kappa1}
\end{eqnarray}
where $\gamma=1+1/n$ and square bracket denotes symmetrization. The following definitions has been done to simplify the equation
\begin{eqnarray}
  \eta_\alpha \equiv \nabla\cdot{\xi}_\alpha,
  \quad
  \eta_{\alpha\beta}\equiv(\nabla_i\xi_\alpha^j)(\nabla_j\xi_\beta^i),
  \quad
  \eta_{\alpha\beta\gamma}\equiv(\nabla_i\xi_\alpha^j)
  (\nabla_j\xi_\beta^k)(\nabla_k\xi_\gamma^i).
\end{eqnarray}
While the first three terms of the integrand in equation (\ref{eq:kappa1}) are responsible for the pressure coupling, the last term governs the gravitational coupling. In the thin disk $p\sim\beta^2\rho$ and the last term is therfore $beta$-times smaller than the previous ones.

\subsection{Isothermal Disk}
We have then looked at the case of an isothermal disk. The equation for the coupling coefficients (equation (\ref{eq:kappa1})) will simplify due to the fact that $\gamma=1$ for isothermal disks and if only terms with the lowest order of $\beta$ are kept. The only surviving term in (\ref{eq:kappa1}) is then
\begin{equation}
  \kappa_{\alpha\beta\gamma} =
  \int_V 
  p(\nabla_i\xi_\alpha^j)(\nabla_j\xi_\beta^k)(\nabla_k\xi_\gamma^i)
  \dd V,
  \label{eq:kappaiso}
\end{equation}

where
\begin{eqnarray}
(\nabla_i\xi_\alpha^j)(\nabla_j\xi_\beta^k)(\nabla_k\xi_\gamma^i)= 
(\partial_i\xi_\alpha^j+\Gamma_{il}^j\xi_\alpha^l) (\partial_j\xi_\beta^k+\Gamma_{jm}^k\xi_\beta^m)
  (\partial_k\xi_\gamma^i+\Gamma_{kn}^i\xi_\gamma^n).
  \label{eq:kappaiso2}
\end{eqnarray}
Because low power of $\beta$ is desired the only term kept from (\ref{eq:kappaiso2}) is 
\begin{equation}
  (\partial_i\xi_\alpha^j)(\partial_j\xi_\beta^k)(\partial_k\xi_\gamma^i).
  \label{kappaiso3}
\end{equation}
If symmetry in $\phi$ direction is assumed and the indicies are run over the scaled r and z, see equation (\ref{eq:rescaling of r and z}), the result will be

\begin{eqnarray}
  (\partial_i\xi_\alpha^j)(\partial_j\xi_\beta^k)(\partial_k\xi_\gamma^i) \!\!\!&=&\!\!\!
   \frac{1}{\beta^3r_0^3} \Big\{
    (\partial_{\bar{x}}\xi_\alpha^r)(\partial_{\bar{x}}\xi_\beta^r)(\partial_{\bar{x}}\xi_\gamma^r)
  + (\partial_{\bar{x}}\xi_\alpha^z)(\partial_{\bar{y}}\xi_\beta^r)(\partial_{\bar{x}}\xi_\gamma^r)
  + (\partial_{\bar{x}}\xi_\alpha^z)(\partial_{\bar{y}}\xi_\beta^z)(\partial_{\bar{y}}\xi_\gamma^r)
  \nonumber\\ \!\!\!&\phantom{=}&\!\!\!
  + (\partial_{\bar{y}}\xi_\alpha^z)(\partial_{\bar{y}}\xi_\beta^r)(\partial_{\bar{x}}\xi_\gamma^z)
  + (\partial_{\bar{y}}\xi_\alpha^z)(\partial_{\bar{y}}\xi_\beta^z)(\partial_{\bar{y}}\xi_\gamma^z)
  + (\partial_{\bar{y}}\xi_\alpha^r)(\partial_{\bar{x}}\xi_\beta^r)(\partial_{\bar{x}}\xi_\gamma^z)
  \nonumber\\ \!\!\!&\phantom{=}&\!\!\!
  + (\partial_{\bar{y}}\xi_\alpha^r)(\partial_{\bar{x}}\xi_\beta^z)(\partial_{\bar{y}}\xi_\gamma^z)
  + (\partial_{\bar{x}}\xi_\alpha^r)(\partial_{\bar{x}}\xi_\beta^z)(\partial_{\bar{y}}\xi_\gamma^r)
    \Big\}.
\label{eq:kappaisoindex}
\end{eqnarray}
Looking at one term at a time and using the Lagrangian displacements from equations (\ref{eq:xi_r})- (\ref{eq:xi_z}) gives 

\begin{eqnarray}
(\partial_{\bar{x}}\xi_\alpha^r)(\partial_{\bar{x}}\xi_\beta^r)(\partial_{\bar{x}}\xi_\gamma^r)
\!\!\!&{=}&\!\!\! 
A \ee^{-i(m_{\alpha\beta\gamma}\phi + \bar{k}_{\alpha\beta\gamma} \bar{x})} H_{n_\alpha} H_{n_\beta} H_{n_\gamma} 
\nonumber\\ 
A \!\!\!&{=}&\!\!\!  \frac{K^3\bar{k}_\alpha^2 \bar{k}_\beta^2 \bar{k}_\gamma^2 \bar{\sigma}_\alpha^2 \bar{\sigma}_\beta^2 \bar{\sigma}_\gamma^2}
  {n_\alpha n_\beta n_\gamma ( \bar{\sigma}_\alpha^2-\bar{\omega}_r^2)(\bar{\sigma}_\beta^2-\bar{\omega}_r^2)(\bar{\sigma}_\gamma^2-\bar{\omega}_r^2)},
  \label{eq:A}
\end{eqnarray}

\begin{eqnarray} 
(\partial_{\bar{x}}\xi_\alpha^z)(\partial_{\bar{y}}\xi_\beta^r)(\partial_{\bar{x}}\xi_\gamma^r)
 \!\!\!&{=}&\!\!\! 
 B \ee^{-i(m_{\alpha\beta\gamma}\phi + \bar{k}_{\alpha\beta\gamma} \bar{x})} H_{n_\gamma} H_{n_{\alpha}-1} H_{n_{\beta}-1} \nonumber\\
B \!\!\!&{=}&\!\!\!  \frac{K^3\bar{k}_\alpha\bar{k}_\beta \bar{k}_\gamma^2\bar{\sigma}_\beta^2\bar{\sigma}_\gamma^2}
  {2 n_\gamma (\bar{\sigma}_\beta^2-\bar{\omega}_r^2)(\bar{\sigma}_\gamma^2-\bar{\omega}_r^2)},
\label{eq:B}
\end{eqnarray}

\begin{eqnarray}
(\partial_{\bar{x}}\xi_\alpha^z)(\partial_{\bar{y}}\xi_\beta^z)(\partial_{\bar{y}}\xi_\gamma^r)
\!\!\!&{=}&\!\!\! 
 C \ee^{-i(m_{\alpha\beta\gamma}\phi + \bar{k}_{\alpha\beta\gamma}\bar{x})} H_{n_{\alpha}-1} H_{n_{\beta}-2} H_{n_{\gamma}-1} \nonumber\\
C \!\!\!&{=}&\!\!\!  \frac{K^3 \bar{k}_\alpha \bar{k}_\gamma \bar{\sigma}_\gamma^2(n_\beta-1)}
   {2 (\bar{\sigma}_\gamma^2-\bar{\omega}_r^2)},
 \label{eq:C}
\end{eqnarray}

\begin{eqnarray}
 (\partial_{\bar{y}}\xi_\alpha^z)(\partial_{\bar{y}}\xi_\beta^r)(\partial_{\bar{x}}\xi_\gamma^z)
 \!\!\!&{=}&\!\!\! 
 D \ee^{-i(m_{\alpha\beta\gamma}\phi + \bar{k}_{\alpha\beta\gamma}\bar{x})} H_{n_{\alpha}-2} H_{n_{\beta}-1} H_{n_{\gamma}} \nonumber\\
D \!\!\!&{=}&\!\!\! \frac{iK^3\bar{k}_\beta \bar{k}_\gamma^2 \bar{\sigma}_\beta^2 \bar{\sigma}_\gamma^2 (n_\alpha-1)}
   {2n_\gamma (\bar{\sigma}_\beta^2-\bar{\omega}_r^2)(\bar{\sigma}_\gamma^2-\bar{\omega}_r^2)},
\label{eq:D}
\end{eqnarray}

\begin{eqnarray}
(\partial_{\bar{y}}\xi_\alpha^z)(\partial_{\bar{y}}\xi_\beta^z)(\partial_{\bar{y}}\xi_\gamma^z)
\!\!\!&{=}&\!\!\! 
 E \ee^{-i(m_{\alpha\beta\gamma}\phi + \bar{k}_{\alpha\beta\gamma}\bar{x})} H_{n_{\alpha}-2} H_{n_{\beta}-2} H_{n_{\gamma}-2} \nonumber\\
 E \!\!\!&{=}&\!\!\! {K^3(n_{\alpha}-1)(n_\beta-1)(n_\gamma-1)},
\label{eq:E}
\end{eqnarray}

\begin{eqnarray}
(\partial_{\bar{y}}\xi_\alpha^r)(\partial_{\bar{x}}\xi_\beta^r)(\partial_{\bar{x}}\xi_\gamma^z)
\!\!\!&{=}&\!\!\! 
F \ee^{-i(m_{\alpha\beta\gamma}\phi + \bar{k}_{\alpha\beta\gamma}\bar{x})} H_{n_{\alpha}-1} H_{n_{\beta}} H_{n_{\gamma}} \nonumber\\
  F \!\!\!&{=}&\!\!\! \frac{K^3\bar{k}_\alpha \bar{k}_\beta^2 \bar{k}_\gamma \bar{\sigma}_{\alpha}^2 \bar{\sigma}_{\beta}^2}
  { n_\beta (\bar{\sigma}_\alpha^2-\bar{\omega}_r^2)(\bar{\sigma}_\beta^2-\bar{\omega}_r^2)},
\label{eq:F}
\end{eqnarray}

\begin{eqnarray}
(\partial_{\bar{y}}\xi_\alpha^r)(\partial_{\bar{x}}\xi_\beta^z)(\partial_{\bar{y}}\xi_\gamma^z)
\!\!\!&{=}&\!\!\! 
G \ee^{-i(m_{\alpha\beta\gamma}\phi + \bar{k}_{\alpha\beta\gamma}\bar{x})} H_{n_{\alpha}-1} H_{n_{\beta}-1} H_{n_{\gamma}-2} \nonumber\\
G \!\!\!&{=}&\!\!\! \frac{K^3 \bar{k}_\alpha \bar{k}_\beta (n_\gamma-1)}{2(\bar{\sigma}_\alpha^2-\bar{\omega}_r^2)},
 \label{eq:G}
\end{eqnarray}

\begin{eqnarray}
(\partial_{\bar{x}}\xi_\alpha^r)(\partial_{\bar{x}}\xi_\beta^z)(\partial_{\bar{y}}\xi_\gamma^r)
\!\!\!&{=}&\!\!\! 
J \ee^{-i(m_{\alpha\beta\gamma}\phi + \bar{k}_{\alpha\beta\gamma}\bar{x})} H_{n_{\alpha}} H_{n_{\beta}-1} H_{n_{\gamma}-1} \nonumber\\
 J \!\!\!&{=}&\!\!\! \frac{K^3 \bar{k}_\alpha^2 \bar{k}_\beta \bar{k}_\gamma \bar{\sigma}_\alpha^2 \bar{\sigma}_\gamma^2}  {2 n_\alpha (\bar{\sigma}_\alpha^2-\bar{\omega}_r^2)(\bar{\sigma}_\gamma^2-\bar{\omega}_r^2)}.
\label{eq:J}
\end{eqnarray}
Here $H^{\prime}_m=mH_{m-1}$ ($m\neq0$ and $H^{\prime}_0=0$), $m_{\alpha\beta\gamma}=m_{\alpha}+m_{\beta}+m_{\gamma}$
and $\bar{k}_{\alpha\beta\gamma}=\bar{k}_{\alpha}+\bar{k}_{\beta}+\bar{k}_{\gamma}$ have been used. Adding the individual terms together gives

\begin{eqnarray}
  \kappa_{\alpha\beta\gamma} \!\!\!&=&\!\!\!
  \frac{1}{\beta^3 r_0^3}
  \int_V
  p(A H_{n_\alpha} H_{n_\beta} H_{n_\gamma} 
  + B H_{n_\gamma} H_{n_{\alpha}-1} H_{n_{\beta}-1}
  + C H_{n_{\alpha}-1} H_{n_{\beta}-2} H_{n_{\gamma}-1} 
    \nonumber\\ \!\!\!&\phantom{=}&\!\!\!
  + D H_{n_{\alpha}-2} H_{n_{\beta}-1} H_{n_{\gamma}}
  + E H_{n_{\alpha}-2} H_{n_{\beta}-2} H_{n_{\gamma}-2} 
  + F H_{n_{\alpha}-1} H_{n_{\beta}} H_{n_{\gamma}}
  \nonumber\\ \!\!\!&\phantom{=}&\!\!\!
  + G H_{n_{\alpha}-1} H_{n_{\beta}-1} H_{n_{\gamma}-2}  
  +J H_{n_{\alpha}} H_{n_{\beta}-1} H_{n_{\gamma}-1})
  \ee^{-i(m_{\alpha\beta\gamma}\phi + \bar{k}_{\alpha\beta\gamma}\bar{x})}
\dd V.
\label{eq:kappaiso4}
\end{eqnarray}
The only $\phi$-dependence is in
$\ee^{-im_{\alpha\beta\gamma}\phi}=\cos(m_{\alpha\beta\gamma}\phi)-i\sin(m_{\alpha\beta\gamma}\phi)$.
The integral is over an even interval, which implies that $m_{\alpha\beta\gamma}=0$
otherwise $\kappa$ is equal to zero.
This also applies to the $\bar{x}$-dependence and gives $\bar{k}_{\alpha\beta\gamma}=0$.
For isothermal disks the pressure is $p=c_{s,0}^2 \rho_0 \ee^{-\bar{y}^2/2}$.
Putting all this together gives

\begin{eqnarray}
  \kappa_{\alpha\beta\gamma} \!\!\!&=&\!\!\!
  \frac{4\pi l c_{s,0}^2 \rho_0}{\beta^3 r_0^3}
  \int_0^{\infty}
  \ee^{-\bar{y}^2/2} (A H_{n_\alpha} H_{n_\beta} H_{n_\gamma}  
  +B  H_{n_\alpha-1} H_{n_{\beta}-1}H_{n_\gamma}
   \nonumber\\ \!\!\!&\phantom{=}&\!\!\!
  + C H_{n_{\alpha}-1} H_{n_{\beta}-2} H_{n_{\gamma}-1} 
  + D H_{n_{\alpha}-2} H_{n_{\beta}-1} H_{n_{\gamma}}
  + E H_{n_{\alpha}-2} H_{n_{\beta}-2} H_{n_{\gamma}-2}  
    \nonumber\\ \!\!\!&\phantom{=}&\!\!\!
  +F H_{n_{\alpha}-1} H_{n_{\beta}} H_{n_{\gamma}}
  + G H_{n_{\alpha}-1} H_{n_{\beta}-1} H_{n_{\gamma}-2}  
  +J H_{n_{\alpha}} H_{n_{\beta}-1} H_{n_{\gamma}-1})
\dd \bar{y}.
\label{eq:kappaiso5}
\end{eqnarray}
To simplify  further we have looked at different properties of Hermite polynomials. The product of two Hermite polynomials can be expressed as
\begin{eqnarray}
	H_i H_j =i! j! \sum^{min(i,j)}_{r=0} \frac{1}{r!(i-r)!(j-r)!} H_{i+j-2r}.
\label{eq:hermprod}
\end{eqnarray} 
Multiplying equation (\ref{eq:hermprod}) with a third Hermite polynomial along with the exponential function that appears in equation (\ref{eq:kappaiso5}) and then integrating yields

\begin{eqnarray}
\int_0^{\infty} 
	H_i H_j H_k \ee^{-\bar{y}^2/2} \dd \bar{y}=
\frac{i! j!}{2} \sum^{min(i,j)}_{r=0} \frac{1}{r!(i-r)!(j-r)!}
	\int_{-\infty}^\infty H_{i+j-2r}H_k \ee^{-\bar{y}^2/2} \dd \bar{y}.
\label{eq:herm2}
\end{eqnarray}	
The orthogonality condition for Hermite polynomials is
\begin{equation}
	\int_{-\infty}^\infty H_i H_j \ee^{-\bar{y}^2/2} \dd \bar{y}=
	j!\sqrt{2\pi}\delta_{ij}.
\end{equation}
Using this in equation (\ref{eq:herm2}) gives
\begin{eqnarray}
\int_0^{\infty} 
	H_i H_j H_k \ee^{-\bar{y}^2/2} \dd \bar{y}=
\sqrt{\frac{\pi}{2}} i!j!k! \sum^{min(i,j)}_{r=0}	\frac{1}{r!(i-r)!(j-r)!} \delta_{i+j-2r,k}.
\label{eq:herm1}
\end{eqnarray}
This gives the condition that	
\begin{eqnarray}
	I_{ijk}= \int_0^{\infty} 
	H_i H_j H_k \ee^{-\bar{y}^2/2} \dd \bar{y}\neq0
\end{eqnarray}
when $i+j-2r=k$, that is $r=(i+j-k)/2$.
Using this equation (\ref{eq:herm1}) becomes
\begin{eqnarray}
	I_{ijk}=\sqrt{\frac{\pi}{2}} \frac{i!j!k!}{\left(\frac{i+j-k}{2}\right)!\left(\frac{i-j+k}{2}\right)!\left(\frac{-i+j+k}{2}\right)!}
\end{eqnarray}
with the constraints that $i+j\geq k$, $i+k\geq j$, $j+k\geq i$ and $(i+j+k)/2\in \textbf{Z}$. Now equation (\ref{eq:kappaiso5}) can be expressed simpler
\begin{eqnarray}
  \kappa_{\alpha\beta\gamma} \!\!\!&=&\!\!\!
  \frac{4\pi l c_{s,0}^2 \rho_0}{\beta^3 r_0^3}
  \left(A I_{{n_\alpha},{n_\beta},{n_\gamma}} 
  + B I_{{n_\alpha-1}, {n_{\beta}-1},{n_\gamma}}
  + C I_{{n_{\alpha}-1},{n_{\beta}-2},{n_{\gamma}-1}}
     + D I_{{n_{\alpha}-2},{n_{\beta}-1},{n_{\gamma}}}
    \right. \nonumber\\ \!\!\!&\phantom{=}&\!\!\! \left.
    + E I_{{n_{\alpha}-2},{n_{\beta}-2},{n_{\gamma}-2}} 
    + F I_{{n_{\alpha}-1},{n_{\beta}}, {n_{\gamma}}}
  + G I_{{n_{\alpha}-1},{n_{\beta}-1},{n_{\gamma}-2}} 
 + J I_{{n_{\alpha}},{n_{\beta}-1}, {n_{\gamma}-1}}\right).
 \label{eq:kappaI}
\end{eqnarray}

\subsection{Purely Vertical Oscillations}
As in the isothermal case we only want terms with low order of $\beta$ in equation (\ref{eq:kappa1}), but here $\gamma \neq 1$. Since $\xi^z$ is the only surviving term we get
\begin{eqnarray}
\eta_{\alpha}\eta_{\beta}\eta_{\gamma} = \eta_{[\alpha}\eta_{\beta\gamma]} = \eta_{\alpha\beta\gamma}=
\frac{K^3}{\beta^3 r_0^3} \ee^{-3im_{\alpha\beta\gamma}\phi} (\partial_{\bar{y}}P_{n_\alpha})(\partial_{\bar{y}}P_{n_\beta})(\partial_{\bar{y}}P_{n_\gamma}).
\label{eq:eta}
\end{eqnarray}
This gives coupling coefficients
\begin{eqnarray}
  \kappa_{\alpha\beta\gamma} \!\!\!&=&\!\!\!
  \frac{1}{2} \int_V 
  (p(\gamma-1)^2 + 3p(\gamma-1) + 2p) \frac{K^3}{\beta^3 r_0^3} \ee^{-3im\phi} (\partial_{\bar{y}}P_{n_\alpha})(\partial_{\bar{y}}P_{n_\beta})(\partial_{\bar{y}}P_{n_\gamma})
  \dd V
  \nonumber\\ \!\!\!&=&\!\!\!
  \frac{1}{2} \frac{K^3}{\beta^3 r_0^3}
  \int_V
  \ee^{-3im\phi} p\gamma(\gamma+1) (\partial_{\bar{y}}P_{n_\alpha})(\partial_{\bar{y}}P_{n_\beta})(\partial_{\bar{y}}P_{n_\gamma})
  \dd V.
\end{eqnarray}
In this case the pressure is $p=p_0 (1-\bar{y}^2)^{n+1}$. For the same reasons as in the isothermal case, $m_{\alpha\beta\gamma}=0$. Using this the final expression will be
\begin{eqnarray}
  \kappa_{\alpha\beta\gamma} \!\!\!&=&\!\!\!
  \frac{K^3}{\beta^3 r_0^3} \pi p_0 \gamma (\gamma +1)
  \int_{-1}^1
  (1-\bar{y}^2)^{n+1} (\partial_{\bar{y}}P_{n_\alpha})(\partial_{\bar{y}}P_{n_\beta})(\partial_{\bar{y}}P_{n_\gamma})
  \dd \bar{y}.
\label{eq:kappajacobi}
\end{eqnarray}

\section{Summary of the Results}
\label{sec6}
In the case of isothermal disk, see equation (\ref{eq:kappaiso5}), the following selection rules yield:\\
\\
$n_i\geq 1$ where $i=\alpha, \beta, \gamma$ otherwise $\kappa$ is undefined \\
$n_\alpha, n_\beta, n_\gamma\geq 1$ gives the following constraints on $I_{ijk}$:
\begin{list}{}{}
\item $(i+j+k)/2\in \textbf{Z}$
\item $i+j\geq k$
\item $i+k\geq j$
\item $j+k\geq i$ 
\end{list}
It can be noted that $\kappa$ in this case is always nonzero. Some examples of coupling coefficients follows:
\begin{list}{}{}
\item $\kappa_{111}= \frac{4\pi l c_{s,0}^2 \rho_0}{\beta^3 r_0^3} F\sqrt{\frac{\pi}{2}}$
\item $\kappa_{112}= \frac{4\pi l c_{s,0}^2 \rho_0}{\beta^3 r_0^3} (2A+G+J)\sqrt{\frac{\pi}{2}}$ 
\item $\kappa_{122}= \frac{4\pi l c_{s,0}^2 \rho_0}{\beta^3 r_0^3} 2F\sqrt{\frac{\pi}{2}}$ 
\item $\kappa_{222}= \frac{4\pi l c_{s,0}^2 \rho_0}{\beta^3 r_0^3} (8A+2B+C+E+G+2J)\sqrt{\frac{\pi}{2}}$ 
\item $\kappa_{223}= \frac{4\pi l c_{s,0}^2 \rho_0}{\beta^3 r_0^3} 6F\sqrt{\frac{\pi}{2}}$ 
\item $\kappa_{233}= \frac{4\pi l c_{s,0}^2 \rho_0}{\beta^3 r_0^3} (36A+6B+2C+E+2G+8J)\sqrt{\frac{\pi}{2}}$ 
\item $\kappa_{333}= \frac{4\pi l c_{s,0}^2 \rho_0}{\beta^3 r_0^3} (6D+36F)\sqrt{\frac{\pi}{2}}$ 
\end{list}
 
In the case of only vertical oscillations, see equation (\ref{eq:kappajacobi}), the selection rules becomes more compact. The only cases when $\kappa$ is nonzero are when $n_\alpha$, $n_\beta$ and $n_\gamma$ is \textit{all odd}, or when \textit{two} of them are \textit{even} and \textit{one} is \textit{odd}.

\section{Discussion and Conclusions} 
In this work we examined nonlinear
interactions between linear modes of the thin accretion disk. Our approach
is analogous to those used in problems of nonlinear stellar pulsation (see
e.g. works of \citet{lynost67}, \citet{frisch78}, \citet{schutz80a}, \citet{schutz80b},   
\citet{sch01} and many others). The
Lagrangian displacement that characterize the resulting oscillations is 
given by equation (\ref{eq:xi_alpha}), the time-dependent coefficients $c_\alpha(t)$ are
solutions of the coupled-anharmonic-oscillators equations (\ref{eq:nlin1}). Hence, a
problem of solving nonlinear partial differential equations has been 
reduced to a problem of finding solutions of many ordinary
differential equations (Schenk et al 2003). In the rest of the paper we
concentrated on the effects of quadratic nonlinearities and gave explicit
formulae for the three-mode coupling coefficients
$\kappa_{\alpha\beta\gamma}$. 

Our key assumption is a completeness of the eigenfunctions of the linear 
normal modes. We note that previous analysis of linear oscillations by
groups of Kato and Wagoner result in such complete set of eigenfunctions.
For simplicity, we consider here an idealized situation of an
isothermal disk. Moreover, another substantial approximation is that the
nonlinear interactions among modes are studied in a small annuli whose
radial extension is of the order of the thickness of the disk. Under these
simplifications the eigenfunctions of the disk are remarkably simple. The
next necessary step is to use the WKBJ approximation to describe the
radial structure of the modes. Preliminary results show that the three
modes interact only locally, close to radii, where a condition
$k_\alpha\pm k_\beta\pm k_\gamma \approx0$ is met ($k_\mu$ are
wavevectors of the modes).

The coupling coefficient characterize energy exchange between modes due to
nonlinear interactions. In our analysis the disk is made of a polytropic
flow with no magnetic field. Hence the only contributions to the coupling
coefficient are pressure and gravity forces. As we demonstrated in
section~\ref{sec5} the contribution of the latter is negligible for a thin
accretion disk. 

In section~\ref{sec6} simple selection rules for the three-mode coupling
coefficients are derived. They are based on the symmetry properties of the
modal eigenfunctions. In principle, a nonlinear interaction among modes is
possible only when  $(n_\alpha + n_\beta + n_\gamma)$ is an even number,
where $n_\mu$ denote quantum numbers that characterize vertical structure
of the modes. The second necessary condition is the ``triangle
inequality'': each quantum number  must not be greater then the sum of the
remaining two. Hence, nonlinear interactions of two oscillation modes (with
quantum numbers $n_\alpha$ and $n_\beta$) may lead to an excitation of
many others modes.

There is an important physical difference between the type of resonance 
studied here and that studied in several recent papers by Kato (see \citet{kato03}, \citet{kato04b} and \citet{kato0709}). 
We deal with a resonance that is excited by a nonlinear three-mode 
coupling of disk oscillatory modes, and Kato consideres a nonlinear 
resonant coupling between disk oscillations and its deformation. In Kato's 
picture the deformation (assumed to be a fixed perturbation on the disks) 
is essential, as it provides the energy for the resonance excitation. Contrary 
to what is studied in the present paper, in Kato's resonance no direct couplings 
between the disk oscillations are considered. 

The mathematical formalism used in this paper may also be used
to study several other issues: linear and nonlinear
viscous damping of the modes, possible excitation mechanisms, or
low-frequency modulations of the high-frequency oscillations proposed
by Abramowicz \& Kluzniak as an underlying mechanism for \citet{psa99} 
correlation. They will be addressed in our future research.
We believe that the formalism may be useful for a future development 
of the general theory of nonlinear diskoseismology. 

\medskip
Research reported here was done in connection with our master diploma work at the Physics Department of G\"{o}teborg University, under supervision of M.A.~Abramowicz. We thank him for all his support and encouragement. We also thank J.~Hor\'ak for introducing us to the mathematics of non-linear coupling and for constant help and guidance. We thank P.~Rebusco aswell for explaining us the multiple scales method. Part of this research was carried out at Nordita in Copenhagen and at the Institute of Astronomy in Prague. We thank these institutions, International Astronomical Union, Copernicus Center in Warsaw
and also G\"{o}teborg University, for their support through several research and travel grants,
in particular the Polish KBN grant N203 009 31/1466.

\end{document}